\documentclass[aps,a4paper,prc, amsfonts, amssymb, amsmath, reprint, showkeys, nofootinbib, twoside, superscriptaddress, standalone]{revtex4-1}
\usepackage{lipsum}
\usepackage{newtxtext,newtxmath}
\usepackage{notoccite}
\usepackage{upgreek}
\usepackage{flushend}
\usepackage{array}
\usepackage{varwidth}
\usepackage[colorinlistoftodos, color=green!40, prependcaption]{todonotes}
\itshape
\usepackage{indentfirst}
\usepackage{pgfplots}
\usepackage{tikz}
\usetikzlibrary{shapes.geometric, arrows.meta, positioning}
\usepackage{multirow}
\graphicspath{{img/},{../img/},{pics/}}
\usepackage{soul}\sethlcolor{yellow}\soulregister{\cite}{7}\soulregister{\ref}{7}

\definecolor{kpgreen}{rgb}{0,0.6,0}

\usepackage{braket}
\usepackage{graphicx}
\usepackage{float}
\usepackage{subfigure}
\usetikzlibrary{fit}
\usetikzlibrary{calc,positioning}
\usepackage{amsmath}
\usepackage{color}
\usepackage{hyperref}
\usepackage{calligra}
\hypersetup{hypertex=true,
    colorlinks=true,
    linkcolor=blue,
    urlcolor=blue,
    anchorcolor=blue,
    citecolor=blue}
\usepackage{longtable}
\usepackage{supertabular}
\usepackage{titlesec}

\titleformat{\subsection}[hang]{\bfseries}{}{0.5em}{}
\titlespacing{\subsection}{0pt}{12pt}{8pt}

\begin{document}

\title{Self-Consistent Microscopic Description of the $^{146}$Sm
Half-Life for Early Solar System Chronology}

\author{Shulin Tang}
\affiliation{School of Physics, Nanjing University of Science and Technology, Nanjing 210094, China}

\author{Yibin Qian}
\affiliation{School of Physics, Nanjing University of Science and Technology, Nanjing 210094, China}

\author{Chong Qi}
\affiliation{Department of Physics, Royal Institute of Technology (KTH), SE-10691 Stockholm, Sweden}

\author{Ramon A. Wyss}
\affiliation{Department of Physics, Royal Institute of Technology (KTH), SE-10691 Stockholm, Sweden}

\author{Roberto J. Liotta}
\affiliation{Department of Physics, Royal Institute of Technology (KTH), SE-10691 Stockholm, Sweden}

\author{Dong Bai}
\affiliation{College of Science, Hohai University, Nanjing 211100, China}

\author{Jianjun He}
\affiliation{Institute of Modern Physics, Fudan University, Shanghai 200433, China}

\author{Tao Wan}
\affiliation{Institute of Modern Physics, Fudan University, Shanghai 200433, China}

\author{Bo Zhou}
\affiliation{Institute of Modern Physics, Fudan University, Shanghai 200433, China}

\author{Chang Xu}
\affiliation{School of Physics, Nanjing University, Nanjing 210093, China}

\author{P\'{e}ter Kanizsai}
\affiliation{Konkoly Observatory, HUN-REN Research Centre for Astronomy and Earth Sciences, Konkoly Thege Mikl\'{o}s \'{u}t 15-17., H-1121, Budapest, Hungary}
\affiliation{Institute of Physics and Astronomy, E\"{o}tv\"{o}s Lor\'{a}nd University, P\'{a}zm\'{a}ny P\'{e}ter s\'{e}t\'{a}ny 1/A, H-1117 Budapest, Hungary}

\author{Maria Lugaro}
\affiliation{Konkoly Observatory, HUN-REN Research Centre for Astronomy and Earth Sciences, Konkoly Thege Mikl\'{o}s \'{u}t 15-17., H-1121, Budapest, Hungary}
\affiliation{School of Physics and Astronomy, Monash University, VIC 3800, Australia}

\author{Zhongzhou Ren}
\affiliation{School of Physics Science and Engineering, Tongji University, Shanghai 200092, China}

\date{\today}

\begin{abstract}
The extinct \textit{p}-process radionuclide $^{146}$Sm, decaying to $^{142}$Nd,
is a key short-lived chronometer for the formation and
early evolution of terrestrial planets and planetesimals.
However, there has been a long-standing discrepancy in both the
experimental measurements and the theoretical evaluations of the
$^{146}$Sm half-life. To address this, a self-consistent microscopic framework for
$\alpha$ decay is developed in this work, in which the $\alpha$ particle is
treated as a four-nucleon quartet.
Its formation amplitude in the nuclear
interior is constructed from a coherent
superposition of multiple pure single-particle configurations weighted by
the pairing occupation amplitudes.
Coupling this shell amplitude to the $\alpha$ cluster that
survives in the low-density surface region yields a
self-consistently determined two-state
(shell~$\oplus$~cluster) preformation weight and effective potential.
The calculated half-lives for the neighboring emitters of
$^{146}$Sm are in good agreement with the measured values.
Cross-validation over the
neighboring emitters reproduces each excluded half-life
without any nucleus-specific adjustment, and yields $^{146}$Sm
predictions consistent to within $\pm 6\%$ of the resulting half-life
$T_{1/2}=75.1^{+6.1}_{-5.6}$~Ma.
This provides a solid calibration for theoretical studies of the early solar system.
\end{abstract}

\maketitle

\textit{Introduction.}---In addition to the stable nuclei,
radionuclides were also present in the early
solar system (ESS). Indeed, the presence of a number of such nuclides has been
inferred from excesses of their daughter nuclides
preserved in meteorites \cite{Dauphas2011,Lugaro2018}.
These nuclei are usually referred to as short-lived radioactive
nuclei and are characterized by half-lives of $\sim$0.1 to 100~Ma.
One such nucleus is the \textit{p}-process radionuclide $^{146}$Sm, which has a half-life of
the order of tens of Ma and is therefore now extinct in
the solar system. This nucleus decays via $\alpha$ emission into
stable $^{142}$Nd, providing a sensitive clock for the
formation and early evolution of terrestrial planets and
planetesimals \cite{Prinzhofer1989,Harper1992,Caro2003,Boyet2005}.
Excesses and
deficits of $^{142}$Nd measured in terrestrial, lunar, Martian, and
meteoritic samples date the earliest episodes of silicate
differentiation \cite{Bennett2007,Rankenburg2006,Debaille2007,Boyet2010}, most directly the
separation of planetary mantles \cite{Caro2008}. The half-life of $^{146}$Sm plays an essential role in
the use of the $^{146}$Sm--$^{142}$Nd chronometer.

In spite of its significance, the experimental measurement of the $^{146}$Sm half-life
remains the subject of a long-standing controversy.
The earliest measurements reported $50$~Ma from $\alpha$-particle counting in nuclear
emulsions \cite{Dunlavey1953} and $74 \pm 15$~Ma from an $\alpha$-activity
measurement \cite{Nurmia1964}.
While the proposal of $^{146}$Sm as a \textit{p}-process chronometer
\cite{Audouze1972} was based on a half-life of $70$~Ma, consistent
with the latter \cite{Nurmia1964}, classical counting
experiments instead converge on $103 \pm 5$~Ma \cite{Friedman1966,Meissner1987}.
This higher value aligns with a cross-calibration for meteoritic and lunar samples including the
Al--Mg-dated achondrite meteorite Erg
Chech 002 ($102 \pm 5$~Ma) \cite{Fang2022}, and another cross-calibration against the Pb--Pb and
$^{147}$Sm--$^{143}$Nd clocks for various meteoritic and lunar rocks also
favored the 103~Ma value \cite{Marks2014}.
Recently, however, two direct decay-counting determinations, one by
$\alpha$ spectrometry and one by cryogenic microcalorimetry,
both fall within a lower range of 84--95~Ma \cite{Chiera2024,Kavner2024,Shollenberger2026}.
Moreover, the $85.8 \pm 0.6$~Ma \cite{Shollenberger2026} and $103 \pm 5$~Ma \cite{Friedman1966,Meissner1987}
values are incompatible even at the $3\sigma$ level, so that
no single half-life seems to be accepted without controversy at present.
In practice, the span among the existing measurements results, e.g., in
differences of up to
100~Myr in the $^{146}$Sm--$^{142}$Nd model ages of magmatic events
that occurred toward the end of the Hadean eon \cite{Fang2022}. The
unsettled decay constant of the $^{146}$Sm--$^{142}$Nd system, i.e.,
the half-life of $^{146}$Sm, thus remains a limiting factor for ESS
cosmochronology \cite{Amelin2025}.

On the theoretical side, predictions of the $^{146}$Sm
half-life scatter widely from model to model, spreading over more than
a factor of six
\cite{XuRen2005,Bao2014,Qian2014,Ghodsi2015,Sun2016,Tavares2023,Tang2025,Wu2026},
so that the theoretical results are likewise highly controversial.
Previous approaches to the $\alpha$ decay of $^{146}$Sm generally lack
a detailed dynamical description of the decay process. Owing to the
complexity of the quantum many-body problem, a detailed treatment of
the $\alpha$ formation probability appears to have been achieved
mainly in approaches that combine shell-model configurations with an
explicit $\alpha$-cluster component
\cite{Varga1992,Varga1992NPA,Lovas1998} and in approaches that
describe the $\alpha$ particle as a four-nucleon quartet in the
nuclear medium \cite{Ropke2014,Xu2016,Yang2020}.
In the present work, a self-consistent framework is developed that
explicitly treats the coherent coupling between the shell and cluster
components of the $\alpha$ formation amplitude in the nuclear surface
region. Specifically, the microscopic formation of the $\alpha$ cluster and its
subsequent penetration of the barrier are treated in a unified
way. The $\alpha$ particle is described as a four-nucleon ($2p+2n$) quartet
in the nuclear medium \cite{Ropke2014,Xu2016,XuC2017}, which emerges
as an $\alpha$ cluster in the low-density region \cite{Delion2020} and
overlaps with the shell configurations in the surface region. The
preformation weight then follows from the coupling of these two
components, rather than being assumed. Both components are constructed
from Hartree--Fock--Bogoliubov (HFB) calculations with a single
energy-density functional, which incorporate the pairing correlations
essential for nucleon clustering
\cite{Tonozuka1979,Andreyev2013,QiAndreyev2010}.

Fortunately, the half-lives of several long-lived $\alpha$ emitters
neighboring $^{146}$Sm have been measured with high precision,
offering a valuable opportunity to further sharpen the theoretical
prediction for $^{146}$Sm by incorporating these measurements as
stringent constraints within the present framework. For this
purpose, the ratios of calculated to experimental half-lives are
employed to refine the predicted half-life of $^{146}$Sm in
conjunction with well-established $\alpha$-decay systematics
\cite{Qi2009,Ni2008,Royer2000}, with the robustness of the result
assessed through leave-one-out (LOO) cross-validation. The resulting
theoretical determination of the $^{146}$Sm half-life, accompanied
by a quantified uncertainty, provides an independent calibration
for $^{146}$Sm--$^{142}$Nd cosmochronology.

\textit{Theoretical framework.}---The quartet wave
function is decomposed into a
center-of-mass (c.o.m.)\ part and an intrinsic part,
$\Psi(\mathbf{R},s_{j})=\Phi(\mathbf{R})\,
\varphi^{\mathrm{intr}}(s_{j},\mathbf{R})$, where $\mathbf{R}$ is the
c.o.m.\ coordinate of the four nucleons and $s_{j}$ denotes the
intrinsic Jacobi coordinates \cite{Ropke2014,Xu2016,Yang2020,Yang2021}. The
c.o.m.\ motion is given by
\begin{multline}
-\frac{\hbar^{2}}{8m}\nabla_{R}^{2}\,\Phi(\mathbf{R})
-\frac{\hbar^{2}}{4m}\int d^{9}s_{j}\,
\varphi^{\mathrm{intr},*}(s_{j},\mathbf{R})
\bigl[\nabla_{R}\varphi^{\mathrm{intr}}(s_{j},\mathbf{R})\bigr]\\
\times\bigl[\nabla_{R}\Phi(\mathbf{R})\bigr]
-\frac{\hbar^{2}}{8m}\int d^{9}s_{j}\,
\varphi^{\mathrm{intr},*}(s_{j},\mathbf{R})
\bigl[\nabla_{R}^{2}\,\varphi^{\mathrm{intr}}(s_{j},\mathbf{R})\bigr]\\
\times\Phi(\mathbf{R})
+\int d^{3}R'\,W(\mathbf{R},\mathbf{R}')\,\Phi(\mathbf{R}')
=E\,\Phi(\mathbf{R}),
\label{eq:com}
\end{multline}
where $-\hbar^{2}\nabla_{R}^{2}/(8m)$ is the kinetic energy of the
c.o.m.\ motion with $m$ the nucleon mass,
$W(\mathbf{R},\mathbf{R}')$ is the effective c.o.m.\ potential, and
$E$ is the quartet energy; the second and third terms couple the
c.o.m.\ and intrinsic motions. These coupling terms, involving
$\nabla_{R}\varphi^{\mathrm{intr}}$, vanish identically in
homogeneous nuclear matter and reduce to gradient corrections to the
local-density treatment in a finite nucleus. In the following, they are therefore
neglected \cite{Ropke2014,Xu2016,Yang2020,Yang2021}. In the
local approximation,
$W(\mathbf{R},\mathbf{R}')=W(R)\,\delta(\mathbf{R}-\mathbf{R}')$, with
the c.o.m.\ motion being a pure $S$ wave for the $0^{+}\!\to\!0^{+}$ decays
considered here. The
intrinsic
motion of
the quartet defines the in-medium intrinsic energy
$\widetilde{W}^{\mathrm{intr}}(\rho_{B},\delta)$
\cite{Yang2020,Yang2021,Wu2026}, where
$\rho_{B}=\rho_{n}+\rho_{p}$ is the daughter baryon density and
$\delta=(\rho_{n}-\rho_{p})/\rho_{B}$ the isospin asymmetry.
Pauli blocking arises from
nucleon
exchange between the quartet and the core and strengthens as their
densities overlap. With increasing density, this blocking shifts the
intrinsic energy away from the free-$\alpha$ binding energy until, at
the Mott density, the quartet dissolves into a four-nucleon
shell-model configuration. The effective potential therefore has two
regions separated by the critical
(Mott) radius $R_{c}$, defined by
$\rho_{B}(R_{c})=\rho_{c}=0.02917$~fm$^{-3}$
\cite{Ropke2014,Xu2016,Yang2020,Yang2021,Wu2026}.

The four quartet nucleons are distributed over the partially
occupied orbitals near the Fermi surface of the parent nucleus.
Accordingly, each of the proton and neutron pairs is described by a
coherent superposition of multiple single-particle configurations,
with amplitudes built from the occupation amplitudes in the canonical basis of the HFB calculations, whose
pairing strengths are adjusted to the empirical odd--even mass
differences \cite{Satula1998}. The c.o.m.\ density of each pair
contains, in addition to the
occupation-weighted diagonal terms, interference contributions between
different configurations, which add constructively because all
amplitudes have the same sign. Such coherent configuration mixing has
been shown in shell-model treatments to enhance the $\alpha$-cluster
formation probability by one to two orders of magnitude relative to a
single valence configuration \cite{Tonozuka1979,Varga1992}. The
convolution of the proton- and neutron-pair densities with respect to
the c.o.m.\ of the quartet yields the quartet density $\rho_{4}$,
which defines the shell amplitude $u_{S}(R)\propto R\sqrt{\rho_{4}(R)}$.
The detailed construction is given in the Supplemental Material (SM)
\cite{SM}\nocite{RingSchuck,Soloviev1962,MangRasmussen1962,Mang1964,Macfarlane1961,AlBataina1987,Gupta1970,Casali2016,Nica2026,Prestwood1981,Fulop2003,MacFarlane1961Dy,Holden1985,Mahunka1966,Golovkov1968,Gono1971,Chiera2022,Nica2025,NUBASE2020,Villa2020}.

Beyond the Mott radius the $\alpha$ cluster survives as a
quasi-bound state, and its spatially extended wave
function
penetrates inward across the repulsive Pauli-blocking potential and
overlaps
with the shell formation amplitude in the surface region. The
problem is thus treated as a two-state (shell~$\oplus$~cluster)
coupling.
The tail of the cluster amplitude $u_{C0}$, solved in
the $\alpha$--core interaction potential outside $R_{c}$,
couples to $u_{S}$, and the resulting mixed formation amplitude
is
\begin{equation}
u_{m}=a\,u_{S}+b\,u_{C0}.
\label{eq:mix}
\end{equation}
Since the inward tail of the cluster state overlaps the shell
amplitude in the surface region (see Fig.~S1 in the SM~\cite{SM}), the two
basis states are not
orthogonal, and the coefficient $b^{2}$ is not by itself a
probability. Following a two-state analog of the Fliessbach
prescription \cite{Fliessbach1976,Varga1992NPA,Lovas1998}, the
basis is symmetrically orthogonalized through the square root of
its overlap matrix $\mathcal{S}$, and the $\alpha$-clustering
content of an amplitude is defined as its squared cluster component
in the orthonormalized basis. Requiring $u_{m}$ to carry the
cluster content $P_{c}$ then fixes
$(a,b)=\mathcal{S}^{-1/2}(\sqrt{1-P_{c}},\sqrt{P_{c}})$.
In the same basis $\{u_{S},u_{C0}\}$, the Hamiltonian defines a
$2\times 2$ generalized eigenvalue problem whose matrix elements are
all computed from the same HFB solution. The cluster content of
the shell-dominant eigenvector is identified with the
$\alpha$-cluster preformation probability
\begin{equation}
P_{\alpha}=\bigl[(\mathcal{S}^{1/2}\mathbf{c})_{C}\bigr]^{2},
\label{eq:pref}
\end{equation}
where $\mathbf{c}=(c_{S},c_{C})$ collects the eigenvector
coefficients. Physically, the cluster content entering the mixed
amplitude and the cluster
weight of the two-state solution measure the same quantity, the
amount of $\alpha$ clustering in the formation amplitude.
Consequently, $P_{c}$ is determined
iteratively for each emitter by requiring the two-state solution to
reproduce the mixed amplitude from which the interior potential is
generated, $\mathbf{c}=(a,b)$, or equivalently $P_{\alpha}=P_{c}$. Details
of the two-state coupling are given in the SM \cite{SM}.

Inside $R_{c}$, the c.o.m.\ effective potential is constructed from the
mixed amplitude itself. In this construction, $u_{m}$ is required to solve the local form of
Eq.~(\ref{eq:com}), and inverting the wave equation for the potential gives
\begin{equation}
W(R)-E=\frac{\hbar^{2}}{8m}\,\frac{u_{m}''(R)}{u_{m}(R)}
=\frac{\hbar^{2}}{8m}\,\frac{a\,u_{S}''(R)+b\,u_{C0}''(R)}
{a\,u_{S}(R)+b\,u_{C0}(R)},
\label{eq:madelung}
\end{equation}
where $E$ is the c.o.m.\ energy of the quartet. The potential $W(R)$ is then joined continuously
with the $\alpha$--core interaction in the surface region at $R_{c}$
\cite{Xu2016,XuC2017,Yang2020}, and this matching fixes $E$. The composite potential
constructed in this way, together with the shell and cluster
amplitudes and the resulting mixed eigenvector, is shown for
$^{146}$Sm in Fig.~S1 of the Supplemental Material \cite{SM}.

Outside $R_{c}$, the direct
effective interaction is the M3Y two-Yukawa form
\cite{Bertsch1977,Satchler1979},
$v(s)=c\,e^{-4s}/4s-d\,e^{-2.5s}/2.5s$,
with $s$ in fm and $(c,d)$ the strengths of the short-range
repulsive and long-range attractive parts, respectively. The
nuclear part $V_{N}$ is obtained by double folding $v(s)$ over the
Gaussian matter density of the $\alpha$ particle and the daughter baryon density; the Coulomb part $V_{C}$
is the corresponding folding of the $\alpha$ charge density against the daughter proton density. The exchange
interaction is
intrinsically nonlocal, depending on the full wave function rather
than solely on the local density \cite{Yang2020,Yang2021}.
Accordingly, the in-medium intrinsic energy
$\widetilde{W}^{\mathrm{intr}}(\rho_{B},\delta)$ is obtained by
solving the intrinsic motion equation in momentum space, as described
in Refs.~\cite{Ropke2014,Wu2026}, with Pauli blocking incorporated by restricting
the nucleon momenta to states above the proton and neutron Fermi
surfaces of the core, the explicit expression of which is given in the SM
\cite{SM}. The outer potential thus reads
$W=V_{N}+V_{C}+\widetilde{W}^{\mathrm{intr}}(\rho_{B},\delta)$,
where $\widetilde{W}^{\mathrm{intr}}$ approaches the free-$\alpha$
binding energy of $-28.3$~MeV as $\rho_{B}\to 0$.
All structural input, the proton and neutron densities of the
daughter and the canonical basis of the parent, derives from the SLy4
Skyrme functional \cite{Chabanat1998}, solved at the spherical HFB
level in coordinate space \cite{Dobaczewski1984} with the code
\textsc{hfbrad} \cite{BennaceurDobaczewski2005}.

The decay width is
$\Gamma=2\,\gamma^{2}(r_{m})\,P_{\mathrm{num}}(E_{\mathrm{tunnel}})$
\cite{Lane1958}, where $\gamma^{2}$ is the reduced width from the
channel amplitude normalized to unit preformation, with the interior
wave function matched continuously to the exterior solution, and
$P_{\mathrm{num}}$ is the numerically integrated exterior
penetrability at the physical tunneling energy
$E_{\mathrm{tunnel}}=Q_{\alpha}-28.3$~MeV. The half-life is
$T_{1/2}=\hbar\ln 2/(P_{\alpha}\Gamma)$.

\textit{The half-life of $^{146}$Sm.}---The self-consistent
framework developed above is now applied to $^{146}$Sm and its six
neighboring $\alpha$ emitters, $^{144}$Nd, $^{148}$Sm, $^{148}$Gd,
$^{150}$Gd, $^{152}$Gd, and $^{154}$Dy. With only two adjustable
parameters, the M3Y strengths $(c,d)$, fitted to the experimental
half-lives of these six emitters, which span fourteen orders of
magnitude \cite{Chiera2023,NNDC}, these half-lives are reproduced
within a factor of 3.3. This deviation is acceptable for a
microscopic model but remains significant for precision chronometry.
Consequently, the directly calculated half-lives are not adopted as
the final predictions.
The well measured half-lives of the six emitters neighboring
$^{146}$Sm are accurately described by the universal decay law
(UDL) \cite{Qi2009,RenRen2012,Akrawy2017,Akrawy2019}, and the fit yields a coefficient of
determination of $R^{2}=0.9999$. The UDL has also been shown to describe
calculated $\alpha$-decay half-lives with an equally good linear
correlation
\cite{QiXu2009,QiLiotta2019}. The ratios of the calculated to the
experimental half-lives are consequently anticipated to exhibit a
similar degree of agreement with the UDL. Therefore, to increase the accuracy and precision of our estimate
of the $^{146}$Sm half-life we correct its directly calculated value
by exploiting these ratios, which are insensitive to the specific
choice of parameters.

\begin{figure}
	\includegraphics[width=\columnwidth]{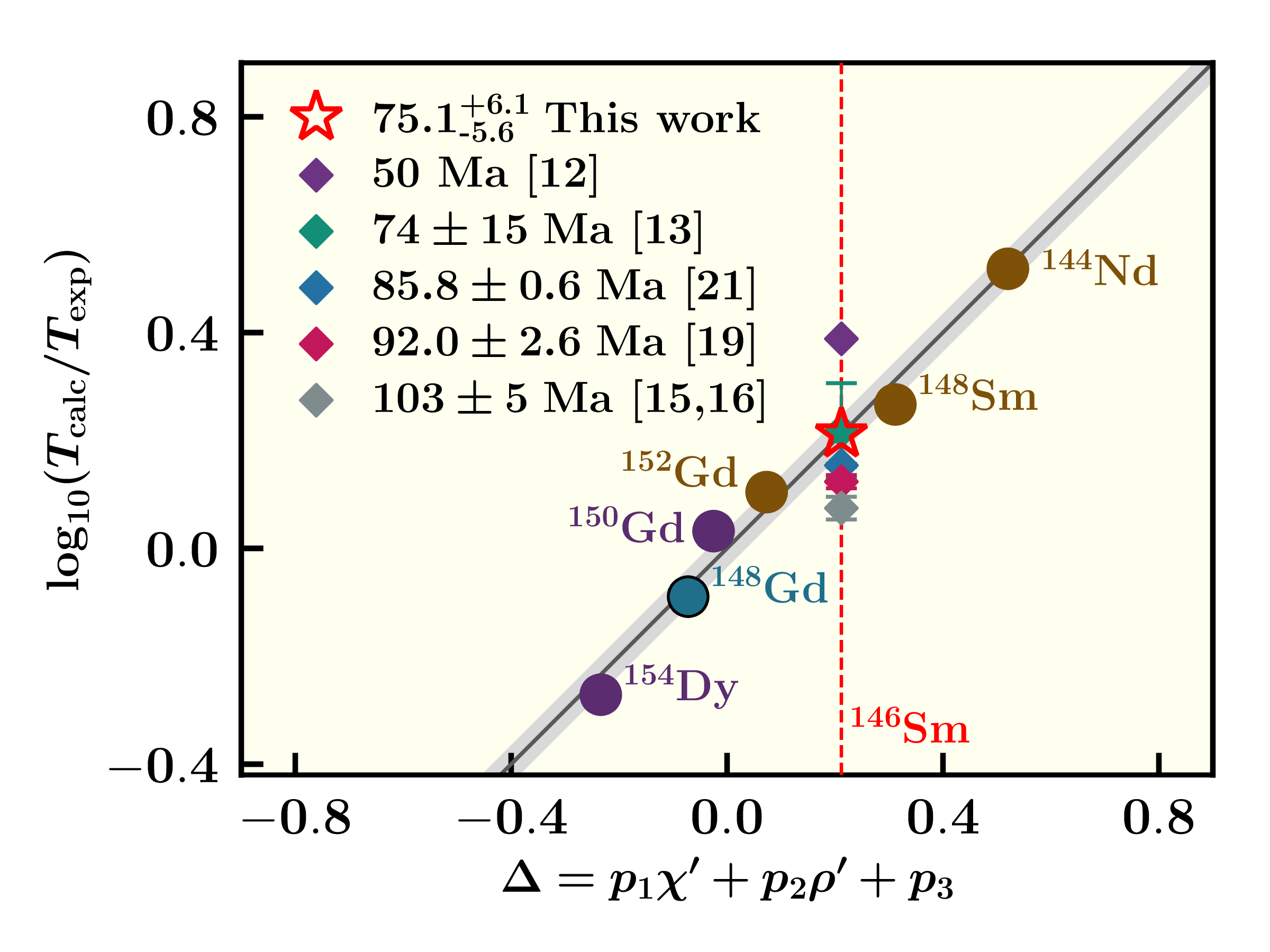}
	\caption{Ratio of the calculated to the experimental half-lives
		versus the UDL residual regression. $^{146}$Sm is not included in
		the fit owing to the lack of an adopted experimental value.
		The solid line marks $\log_{10}(T_{\mathrm{calc}}/T_{\mathrm{exp}})=\Delta$,
		and the gray band indicates an rms deviation of 0.034.}
	\label{fig:udl}
\end{figure}

As shown in Fig.~\ref{fig:udl}, these ratios follow the UDL. The UDL
combination reads $\Delta=p_{1}\chi'+p_{2}\rho'+p_{3}$, with
$\chi'=Z_{\alpha}Z_{d}\sqrt{\bar{A}/Q_{\alpha}}$ and
$\rho'=\sqrt{\bar{A}Z_{\alpha}Z_{d}(A_{\alpha}^{1/3}+A_{d}^{1/3})}$,
where $\bar{A}=4A_{d}/(4+A_{d})$ is the reduced mass number. The
coefficients $p_{1}$, $p_{2}$, and $p_{3}$ are determined by a
least-squares
regression of the residuals of the six neighboring emitters.
The half-life of $^{146}$Sm, obtained by dividing the directly
calculated half-life by $10^{\Delta}$, is $75.1^{+6.1}_{-5.6}$~Ma.
Furthermore, the same residual regression is repeated with two other
widely used empirical formulas, the NRDX form \cite{Ni2008},
$\Delta=p_{1}\sqrt{\bar{A}}\,Z_{\alpha}Z_{d}/\sqrt{Q_{\alpha}}
+p_{2}\sqrt{\bar{A}Z_{\alpha}Z_{d}}+p_{3}$, and the Royer form
\cite{Royer2000},
$\Delta=p_{1}\,A^{1/6}\sqrt{Z}+p_{2}\,Z/\sqrt{Q_{\alpha}}+p_{3}$,
where $A$ and $Z$ are the mass and proton numbers of the parent
nucleus. The calculated half-lives of $^{146}$Sm within these two
forms are $74.8^{+5.9}_{-5.4}$~Ma and $75.1^{+6.2}_{-5.6}$~Ma,
respectively. This consistency demonstrates the robustness of
the present framework.

\begin{figure}
\includegraphics[width=\columnwidth]{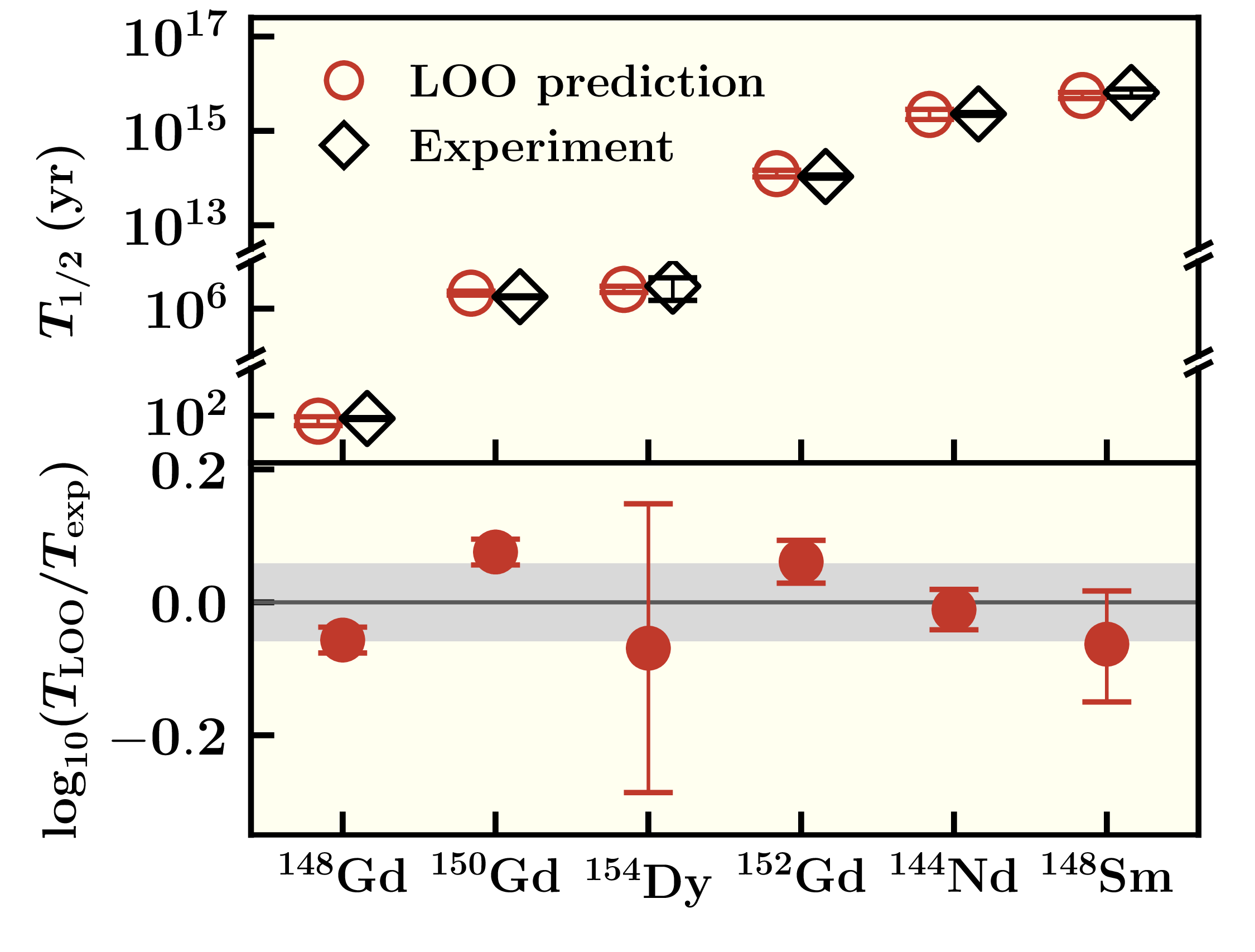}
\caption{Leave-one-out validation of the calibration on the six
neighboring emitters. The top panel compares the LOO-calculated
half-lives with $1\sigma$ uncertainties (red circles) and the
experimental values (black diamonds), while the bottom panel shows the
ratios of calculated to experimental half-lives.}
\label{fig:loo}
\end{figure}

In the present context, we apply leave-one-out cross-validation to
test the insensitivity of the prediction to the fitted
$(c,d)$ parameter set. Concretely, each of the six neighboring
emitters is left out in turn and predicted from the remaining five,
with the shared $(c,d)$ pair and the UDL residual regression refit
in every fold, and its half-life is then
calculated in the same manner as in Fig.~\ref{fig:udl}. As shown in
Fig.~\ref{fig:loo}, the predicted half-lives of all six
emitters agree with the measured values within a factor of 1.2. The calculated and experimental half-lives are
discussed further in the SM \cite{SM}.
The fitted $(c,d)$ pair differs from one LOO calculation to another,
whereas the resulting $^{146}$Sm half-life remains consistent, which
underlines the independence of the prediction from the specific
parameter set. Importantly, the $^{146}$Sm center obtained across
the six independent LOO calculations varies only between 70.7 and
78.2~Ma. Adopting the LOO rms for the uncertainty propagation leaves the
central value unchanged and enlarges the uncertainty only from about
8\% to 14\%. In this
context, the final calculated half-life of $^{146}$Sm and its
associated uncertainty are determined by the six neighboring
emitters, given that their half-lives are experimentally measured.

To assess the sensitivity to the functional, the calculations are
repeated with the Skyrme parametrizations SkM* \cite{Bartel1982},
SkP \cite{Dobaczewski1984}, and SLy7 \cite{Chabanat1998}. The
calculated half-lives of $^{146}$Sm and of the six neighboring
emitters vary within 10\%; the largest spread, between SkM* and SLy7,
changes the central $^{146}$Sm value by less than 6\%.

\begin{figure}
\includegraphics[width=\columnwidth]{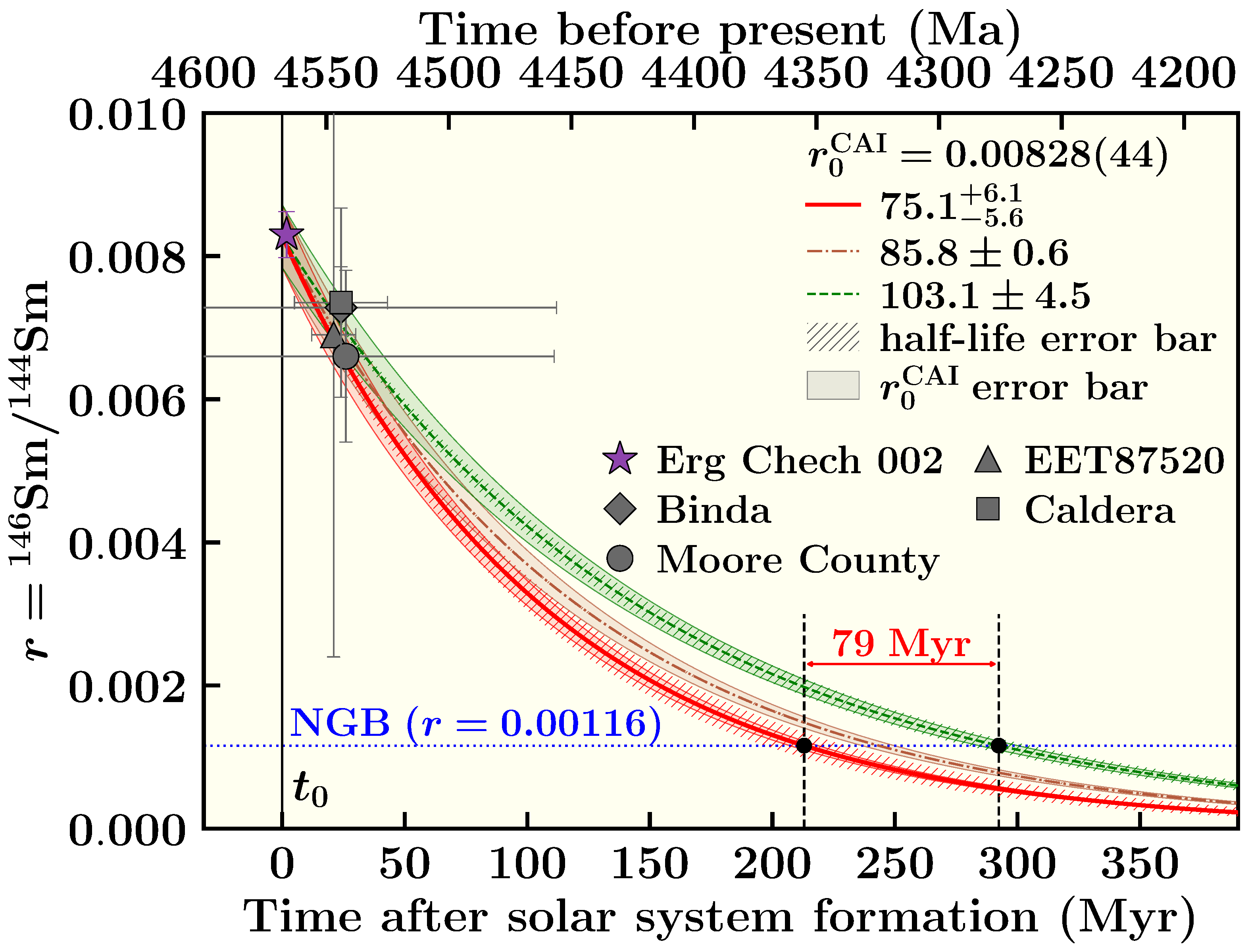}\vspace{-1pt}
\caption{Evolution of the solar system ratio
$r={}^{146}$Sm/$^{144}$Sm from the CAI initial ratio
$r_{0}^{\mathrm{CAI}}=0.00828(44)$ \cite{Marks2014} for the present
half-life (red curve) compared with the half-lives of
85.8~Ma \cite{Shollenberger2026} (brown curve) and
103.1~Ma \cite{Meissner1987} (green curve);
the shaded bands show the $2\sigma$ uncertainty of
$r_{0}^{\mathrm{CAI}}$ and the hatched bands the $1\sigma$ uncertainty
of the respective half-life. The star denotes Erg Chech 002
\cite{Fang2022}, plotted at its Al--Mg crystallization age. The other symbols denote the four
eucrites with concordant $^{147}$Sm--$^{143}$Nd and
$^{146}$Sm--$^{142}$Nd systematics selected in
Ref.~\cite{Boyet2010}, plotted at their $^{147}$Sm--$^{143}$Nd
isochron ages, which are independent of the $^{146}$Sm
half-life. The blue line marks the
$^{146}$Sm/$^{144}$Sm abundance of the enriched NGB source
\cite{ONeil2008}, and the black dots denote its formation times for 75.1 and
103.1~Ma, which differ by 79~Myr.}
\label{fig:chrono}
\end{figure}

\textit{Implications for $^{146}$Sm--$^{142}$Nd
cosmochronology.}---Any chronological application of the
$^{146}$Sm--$^{142}$Nd system rests on the solar system initial
ratio $r_{0}=(^{146}\mathrm{Sm}/^{144}\mathrm{Sm})_{0}$,
conventionally referenced to the time of solar system formation
($t_{0}=4568$~Ma \cite{Bouvier2010}). The initial ratio obtained from the
$^{146}$Sm--$^{142}$Nd isochron of an Allende calcium--aluminum-rich
inclusion (CAI) \cite{Marks2014}, $r_{0}^{\mathrm{CAI}}=0.00828(44)$ ($2\sigma$),
is adopted here as the starting point of the decay curves in
Fig.~\ref{fig:chrono}. This ratio is measured in the
oldest solids of the solar system, whose formation corresponds to
$t_{0}$ by definition, so that it is independent of the half-life,
unlike values back-corrected from samples whose Sm--Nd system closed
some time after $t_{0}$.
An independent check is provided by the achondrite Erg Chech 002,
which crystallized 1.80(1)~Myr after the CAIs and records
$^{146}\mathrm{Sm}/^{144}\mathrm{Sm}=0.00830(32)$ \cite{Fang2022}.
Over this short interval the back-correction is nearly independent
of the half-life, giving $r_{0}=0.00844(33)$, $0.00842(32)$, and
$0.00840(32)$ for 75.1, 85.8, and 103.1~Ma, respectively. All three
values are consistent with the CAI value within their
uncertainties. The four eucrites of Ref.~\cite{Boyet2010} compared with the
decay curves in Fig.~\ref{fig:chrono} carry large error bars of up
to 88~Myr in their $^{147}$Sm--$^{143}$Nd ages. Moreover, all four eucrites lie within $1\sigma$ of each of
the three curves, indicating that they are insufficient to constrain
the half-life. More precise meteoritic data are therefore needed in
future studies.
In contrast, cross-calibrations of the $^{146}$Sm--$^{142}$Nd system
against other chronometers favor a longer half-life,
$102\pm5$~Ma \cite{Fang2022} and about 103~Ma
\cite{Marks2014}, whereas the recent direct decay-counting measurements
\cite{Chiera2024,Kavner2024,Shollenberger2026} and the present
calculation give shorter values. A recalibration of the chronometer
is therefore not straightforward. The chronometric
relations are given in the Supplemental Material \cite{SM}.

On the other hand, dating a planetary-scale silicate
differentiation event requires both the initial
$^{146}$Sm/$^{144}$Sm ratio and the $^{146}$Sm half-life. The
$^{146}$Sm/$^{144}$Sm ratio preserved in the differentiated source
region at the time of its formation can then be converted, through
the radioactive decay law governed by the $^{146}$Sm half-life,
into the elapsed time since $t_{0}$. The
$^{146}$Sm--$^{142}$Nd chronometer is therefore highly
sensitive to the adopted $^{146}$Sm half-life. As
mentioned above, owing to the inconsistency of the measured
half-lives, cosmochronological studies have had to report the
ages of early differentiation events for alternative half-lives
\cite{Fang2022}, a duplication that impedes the
potential of $^{146}$Sm--$^{142}$Nd systematics for the early
solar system chronology. As illustrated in
Fig.~\ref{fig:chrono}, the enriched source of
the Nuvvuagittuq greenstone belt (NGB), a terrestrial crustal
remnant in northern Qu\'{e}bec, Canada \cite{ONeil2008},
formed $212.9^{+17.3}_{-15.9}$~Myr after $t_{0}$ (4355~Ma) for
the present half-life, compared with 243.3 and 292.3~Myr for
85.8~Ma \cite{Shollenberger2026} and 103.1~Ma, i.e., 30 and 79~Myr
later, respectively. For the Moon, dating the solidification of the lunar magma
ocean, the lunar array \cite{Caro2011} gives
$173.9^{+14.1}_{-13.0}$~Myr and the six lunar basalts
\cite{Rankenburg2006} give $167.7^{+13.6}_{-12.5}$~Myr, compared
with $238.7\pm10.4$ and $230.3\pm10.1$~Myr for 103.1~Ma. Note that the shorter
half-life was shown to provide ages in better agreement with lunar
chronology \cite{Shollenberger2026}. For Mars, the source of the
enriched shergottites \cite{Debaille2007} gives
$84.1^{+6.8}_{-6.3}$~Myr and the combined shergottite suite
\cite{Borg2016} gives $45.9^{+3.7}_{-3.4}$~Myr, compared with $115.4\pm5.0$ and $63.1\pm2.8$~Myr for
103.1~Ma.
All half-life and age uncertainties quoted here are $1\sigma$.
The uncertainties of the
measured isotopic data (meteoritic, lunar, and terrestrial samples),
reported as $2\sigma$ (95\% confidence) limits in the original
studies, are not propagated into these ages. It should be noted that the contribution of the $r_{0}^{\mathrm{CAI}}$
uncertainty to the inferred formation ages remains constant for a given
$^{146}$Sm half-life, whereas that of the half-life grows in proportion to the
elapsed time. Since the chronometer becomes increasingly sensitive
to the half-life for lower $^{146}$Sm/$^{144}$Sm ratios, i.e., for
later events, a precisely determined half-life is essential for dating such
events.

\textit{Summary.}---A self-consistent microscopic framework for
$\alpha$ decay has been established by combining a four-nucleon
quartet description of the $\alpha$ particle with
Hartree--Fock--Bogoliubov
calculations. Constructed from a coherent superposition of single-particle
configurations weighted by the HFB pairing occupation amplitudes,
the quartet formation amplitude in the nuclear interior yields a
self-consistently determined
two-state (shell~$\oplus$~cluster) preformation weight and
effective potential. Within the present framework, the calculated half-lives agree
with the experimental values within a factor of 3.3 across
fourteen orders of magnitude. After regression of the calculated-to-experimental half-life
ratios against the corresponding decay laws, the corrected
half-lives reproduce the measured values to within a factor of 1.2
under leave-one-out cross-validation. The resulting
half-life of $^{146}$Sm, $T_{1/2}=75.1^{+6.1}_{-5.6}$~Ma, is
determined in a parameter-insensitive manner. It remains stable at
the $\pm 6\%$ level under leave-one-out cross-validation and is
insensitive to both the choice of the regression form
and the choice of the Skyrme parametrization of the underlying
energy-density functional.
The uncertainties of the $^{146}$Sm/$^{144}$Sm ratios of the CAI,
Erg Chech 002, and the four eucrites
are too large to constrain the half-life, and a recalibration of the
chronometer must take into account the discrepancies among the existing
determinations. Relative to the commonly used 103~Ma
value, our new determination would shorten the time of early
planetary silicate differentiation by roughly 79, 63--65, and
17--31~Myr in the case of Earth, Moon, and Mars, respectively, with
implications for the chronology of the early solar system.

\textit{Acknowledgments.}---We thank Ulrich Ott for pointing out
the very recent half-life measurement \cite{Shollenberger2026} and for
helpful discussions. This work is supported by the
National Natural Science Foundation of China (Grants
No.~12675157 and No.~12535009) and the Lend\"ulet Program LP2023-10 of
the Hungarian Academy of Sciences.

\bibliography{manuscript}

@article{Boyet2005,
  author  = {Boyet, M. and Carlson, R. W.},
  title   = {$^{142}$Nd evidence for early ({$>$}4.53 {Ga}) global differentiation of the silicate {Earth}},
  journal = {Science},
  volume  = {309},
  pages   = {576},
  year    = {2005},
  doi     = {10.1126/science.1113634}
}

@article{Caro2008,
  author  = {Caro, G. and Bourdon, B. and Halliday, A. N. and Quitt{\'e}, G.},
  title   = {Super-chondritic {Sm/Nd} ratios in {Mars}, the {Earth} and the {Moon}},
  journal = {Nature},
  volume  = {452},
  pages   = {336},
  year    = {2008},
  doi     = {10.1038/nature06760}
}

@article{Friedman1966,
  author  = {Friedman, A. M. and Milsted, J. and Metta, D. and Henderson, D. and Lerner, J. and Harkness, A. L. and Rokop, D. J.},
  title   = {Alpha decay half lives of $^{148}${Gd}, $^{150}${Gd} and $^{146}${Sm}},
  journal = {Radiochim. Acta},
  volume  = {5},
  pages   = {192},
  year    = {1966},
  doi     = {10.1524/ract.1966.5.4.192}
}

@article{Meissner1987,
  author  = {Meissner, F. and Schmidt-Ott, W.-D. and Ziegeler, L.},
  title   = {Half-life and $\alpha$-ray energy of $^{146}${Sm}},
  journal = {Z. Phys. A},
  volume  = {327},
  pages   = {171},
  year    = {1987},
  doi     = {10.1007/BF01292406}
}

@phdthesis{Kavner2024,
  author  = {Kavner, A. R. L.},
  title   = {Measurement of $^{146}${Sm} half-life and ionization produced by 254~{eV}$_{\mathrm{nr}}$ nuclear recoils in germanium},
  school  = {University of Michigan},
  year    = {2024}
}

@article{Chiera2024,
  author  = {Chiera, N. M. and Sprung, P. and Amelin, Y. and Dressler, R. and Schumann, D. and Talip, Z.},
  title   = {The $^{146}${Sm} half-life re-measured: consolidating the chronometer for events in the early {Solar System}},
  journal = {Sci. Rep.},
  volume  = {14},
  pages   = {17436},
  year    = {2024},
  doi     = {10.1038/s41598-024-64104-6}
}

@article{Fang2022,
  author  = {Fang, L. and Frossard, P. and Boyet, M. and Bouvier, A. and Barrat, J.-A. and Chaussidon, M. and Moynier, F.},
  title   = {Half-life and initial {Solar System} abundance of $^{146}${Sm} determined from the oldest andesitic meteorite},
  journal = {Proc. Natl. Acad. Sci. U.S.A.},
  volume  = {119},
  pages   = {e2120933119},
  year    = {2022},
  doi     = {10.1073/pnas.2120933119}
}

@article{Villa2020,
  author  = {Villa, I. M. and Holden, N. E. and Possolo, A. and Ickert, R. B. and Hibbert, D. B. and Renne, P. R.},
  title   = {{IUPAC-IUGS} recommendation on the half-lives of $^{147}${Sm} and $^{146}${Sm}},
  journal = {Geochim. Cosmochim. Acta},
  volume  = {285},
  pages   = {70},
  year    = {2020},
  doi     = {10.1016/j.gca.2020.06.022}
}

@article{Qi2009,
  author  = {Qi, C. and Xu, F. R. and Liotta, R. J. and Wyss, R.},
  title   = {Universal decay law in charged-particle emission and exotic cluster radioactivity},
  journal = {Phys. Rev. Lett.},
  volume  = {103},
  pages   = {072501},
  year    = {2009},
  doi     = {10.1103/PhysRevLett.103.072501}
}

@article{Ropke2014,
  author  = {R{\"o}pke, G. and Schuck, P. and Funaki, Y. and Horiuchi, H. and Ren, Z. and Tohsaki, A. and Xu, C. and Yamada, T. and Zhou, B.},
  title   = {Nuclear clusters bound to doubly magic nuclei: {The} case of $^{212}${Po}},
  journal = {Phys. Rev. C},
  volume  = {90},
  pages   = {034304},
  year    = {2014},
  doi     = {10.1103/PhysRevC.90.034304}
}

@article{Xu2016,
  author  = {Xu, C. and Ren, Z. and R{\"o}pke, G. and Schuck, P. and Funaki, Y. and Horiuchi, H. and Tohsaki, A. and Yamada, T. and Zhou, B.},
  title   = {$\alpha$-decay width of $^{212}${Po} from a quartetting wave function approach},
  journal = {Phys. Rev. C},
  volume  = {93},
  pages   = {011306},
  year    = {2016},
  doi     = {10.1103/PhysRevC.93.011306}
}

@article{Yang2020,
  author  = {Yang, S. and Xu, C. and R{\"o}pke, G. and Schuck, P. and Ren, Z. and Funaki, Y. and Horiuchi, H. and Tohsaki, A. and Yamada, T. and Zhou, B.},
  title   = {$\alpha$ decay to a doubly magic core in the quartetting wave function approach},
  journal = {Phys. Rev. C},
  volume  = {101},
  pages   = {024316},
  year    = {2020},
  doi     = {10.1103/PhysRevC.101.024316}
}

@article{Yang2021,
  author  = {Yang, S. and Xu, C. and R{\"o}pke, G.},
  title   = {$\alpha$-cluster formation and decay: The role of shell structure},
  journal = {Phys. Rev. C},
  volume  = {104},
  pages   = {034302},
  year    = {2021},
  doi     = {10.1103/PhysRevC.104.034302}
}

@article{Bertsch1977,
  author  = {Bertsch, G. and Borysowicz, J. and McManus, H. and Love, W. G.},
  title   = {Interactions for inelastic scattering derived from realistic potentials},
  journal = {Nucl. Phys. A},
  volume  = {284},
  pages   = {399},
  year    = {1977},
  doi     = {10.1016/0375-9474(77)90392-X}
}

@article{Satchler1979,
  author  = {Satchler, G. R. and Love, W. G.},
  title   = {Folding model potentials from realistic interactions for heavy-ion scattering},
  journal = {Phys. Rep.},
  volume  = {55},
  pages   = {183},
  year    = {1979},
  doi     = {10.1016/0370-1573(79)90081-4}
}

@misc{NNDC,
  howpublished = {\url{https://www.nndc.bnl.gov/nudat3/}}
}

@article{NUBASE2020,
  author  = {Kondev, F. G. and Wang, M. and Huang, W. J. and Naimi, S. and Audi, G.},
  title   = {The {NUBASE2020} evaluation of nuclear physics properties},
  journal = {Chin. Phys. C},
  volume  = {45},
  pages   = {030001},
  year    = {2021},
  doi     = {10.1088/1674-1137/abddae}
}

@article{Chiera2023,
  author  = {Chiera, N. M. and Dressler, R. and Sprung, P. and Talip, Z. and Schumann, D.},
  title   = {Determination of the half-life of gadolinium-148},
  journal = {Appl. Radiat. Isot.},
  volume  = {194},
  pages   = {110708},
  year    = {2023},
  doi     = {10.1016/j.apradiso.2023.110708}
}

@article{Prinzhofer1989,
  author  = {Prinzhofer, A. and Papanastassiou, D. A. and Wasserburg, G. J.},
  title   = {The presence of $^{146}${Sm} in the early solar system and implications for its nucleosynthesis},
  journal = {Astrophys. J.},
  volume  = {344},
  pages   = {L81},
  year    = {1989},
  doi     = {10.1086/185536}
}

@article{Harper1992,
  author  = {Harper, C. L. and Jacobsen, S. B.},
  title   = {Evidence from coupled $^{147}${Sm}--$^{143}${Nd} and $^{146}${Sm}--$^{142}${Nd} systematics for very early (4.5-{Gyr}) differentiation of the {Earth}'s mantle},
  journal = {Nature},
  volume  = {360},
  pages   = {728},
  year    = {1992},
  doi     = {10.1038/360728a0}
}

@article{Caro2003,
  author  = {Caro, G. and Bourdon, B. and Birck, J.-L. and Moorbath, S.},
  title   = {$^{146}${Sm}--$^{142}${Nd} evidence from {Isua} metamorphosed sediments for early differentiation of the {Earth}'s mantle},
  journal = {Nature},
  volume  = {423},
  pages   = {428},
  year    = {2003},
  doi     = {10.1038/nature01668}
}

@article{Dauphas2011,
  author  = {Dauphas, N. and Chaussidon, M.},
  title   = {A perspective from extinct radionuclides on a young stellar object: {The} {Sun} and its accretion disk},
  journal = {Annu. Rev. Earth Planet. Sci.},
  volume  = {39},
  pages   = {351},
  year    = {2011},
  doi     = {10.1146/annurev-earth-040610-133428}
}

@article{Lugaro2018,
  author  = {Lugaro, M. and Ott, U. and Kereszturi, {\'A}.},
  title   = {Radioactive nuclei from cosmochronology to habitability},
  journal = {Prog. Part. Nucl. Phys.},
  volume  = {102},
  pages   = {1},
  year    = {2018},
  doi     = {10.1016/j.ppnp.2018.05.002}
}

@article{Bennett2007,
  author  = {Bennett, V. C. and Brandon, A. D. and Nutman, A. P.},
  title   = {Coupled $^{142}${Nd}--$^{143}${Nd} isotopic evidence for {Hadean} mantle dynamics},
  journal = {Science},
  volume  = {318},
  pages   = {1907},
  year    = {2007},
  doi     = {10.1126/science.1145928}
}

@article{Debaille2007,
  author  = {Debaille, V. and Brandon, A. D. and Yin, Q. Z. and Jacobsen, B.},
  title   = {Coupled $^{142}${Nd}--$^{143}${Nd} evidence for a protracted magma ocean in {Mars}},
  journal = {Nature},
  volume  = {450},
  pages   = {525},
  year    = {2007},
  doi     = {10.1038/nature06317}
}

@article{XuRen2005,
  author  = {Xu, C. and Ren, Z.},
  title   = {Favored $\alpha$-decays of medium mass nuclei in density-dependent cluster model},
  journal = {Nucl. Phys. A},
  volume  = {760},
  pages   = {303},
  year    = {2005},
  doi     = {10.1016/j.nuclphysa.2005.06.011}
}

@article{Bao2014,
  author  = {Bao, X. and Zhang, H. and Zhang, H. and Royer, G. and Li, J.},
  title   = {Systematical calculation of $\alpha$ decay half-lives with a generalized liquid drop model},
  journal = {Nucl. Phys. A},
  volume  = {921},
  pages   = {85},
  year    = {2014}
}

@article{Qian2014,
  author  = {Qian, Y. and Ren, Z.},
  title   = {Half-lives of $\alpha$ decay from natural nuclides and from superheavy elements},
  journal = {Phys. Lett. B},
  volume  = {738},
  pages   = {87},
  year    = {2014},
  doi     = {10.1016/j.physletb.2014.09.024}
}

@article{Ghodsi2015,
  author  = {Ghodsi, O. N. and Motevalli, S. M. and Gholami, E.},
  title   = {Half-life of spherical $\alpha$ emitters and intrinsic properties of nuclei},
  journal = {Phys. Rev. C},
  volume  = {91},
  pages   = {034611},
  year    = {2015},
  doi     = {10.1103/PhysRevC.91.034611}
}

@article{Sun2016,
  author  = {Sun, X.-D. and Guo, P. and Li, X.-H.},
  title   = {Systematic study of $\alpha$ decay half-lives for even-even nuclei within a two-potential approach},
  journal = {Phys. Rev. C},
  volume  = {93},
  pages   = {034316},
  year    = {2016},
  doi     = {10.1103/PhysRevC.93.034316}
}

@article{Tavares2023,
  author  = {Tavares, O. A. P. and Terranova, M. L.},
  title   = {A semiempirical evaluation of half-life of $^{146}${Sm} isotope},
  journal = {Mod. Phys. Lett. A},
  volume  = {38},
  pages   = {2350115},
  year    = {2023},
  doi     = {10.1142/S0217732323501158}
}

@article{Wu2026,
  author  = {Wu, Y. and Li, R. and Xu, C.},
  title   = {Calibration of the $\alpha$-decay half-life of $^{146}${Sm} for the chronology of early solar system},
  journal = {Phys. Lett. B},
  volume  = {872},
  pages   = {140113},
  year    = {2026},
  doi     = {10.1016/j.physletb.2025.140113}
}

@article{Tonozuka1979,
  author  = {Tonozuka, I. and Arima, A.},
  title   = {Surface $\alpha$-clustering and $\alpha$-decays of $^{212}${Po}},
  journal = {Nucl. Phys. A},
  volume  = {323},
  pages   = {45},
  year    = {1979},
  doi     = {10.1016/0375-9474(79)90415-9}
}

@article{Varga1992,
  author  = {Varga, K. and Lovas, R. G. and Liotta, R. J.},
  title   = {Absolute alpha decay width of $^{212}${Po} in a combined shell and cluster model},
  journal = {Phys. Rev. Lett.},
  volume  = {69},
  pages   = {37},
  year    = {1992},
  doi     = {10.1103/PhysRevLett.69.37}
}

@article{Delion2020,
  author  = {Delion, D. S. and Dumitrescu, A.},
  title   = {Realistic analytical approach for $\alpha$ decay and clustering},
  journal = {Phys. Rev. C},
  volume  = {102},
  pages   = {014327},
  year    = {2020},
  doi     = {10.1103/PhysRevC.102.014327}
}

@article{XuC2017,
  author  = {Xu, C. and R{\"o}pke, G. and Schuck, P. and Ren, Z. and Funaki, Y. and Horiuchi, H. and Tohsaki, A. and Yamada, T. and Zhou, B.},
  title   = {$\alpha$-cluster formation and decay in the quartetting wave function approach},
  journal = {Phys. Rev. C},
  volume  = {95},
  pages   = {061306},
  year    = {2017},
  doi     = {10.1103/PhysRevC.95.061306}
}

@article{Fliessbach1976,
  author  = {Fliessbach, T. and Mang, H. J.},
  title   = {On absolute values of $\alpha$-decay rates},
  journal = {Nucl. Phys. A},
  volume  = {263},
  pages   = {75},
  year    = {1976},
  doi     = {10.1016/0375-9474(76)90184-6}
}

@article{Varga1992NPA,
  author  = {Varga, K. and Lovas, R. G. and Liotta, R. J.},
  title   = {Cluster-configuration shell model for alpha decay},
  journal = {Nucl. Phys. A},
  volume  = {550},
  pages   = {421},
  year    = {1992},
  doi     = {10.1016/0375-9474(92)90017-E}
}

@article{Lovas1998,
  author  = {Lovas, R. G. and Liotta, R. J. and Insolia, A. and Varga, K. and Delion, D. S.},
  title   = {Microscopic theory of cluster radioactivity},
  journal = {Phys. Rep.},
  volume  = {294},
  pages   = {265},
  year    = {1998},
  doi     = {10.1016/S0370-1573(97)00049-5}
}

@article{Dunlavey1953,
  author  = {Dunlavey, D. C. and Seaborg, G. T.},
  title   = {Alpha activity of $^{146}${Sm} as detected with nuclear emulsions},
  journal = {Phys. Rev.},
  volume  = {92},
  pages   = {206},
  year    = {1953},
  doi     = {10.1103/PhysRev.92.206}
}

@article{Audouze1972,
  author  = {Audouze, J. and Schramm, D. N.},
  title   = {$^{146}${Sm}: {A} chronometer for $p$-process nucleosynthesis},
  journal = {Nature},
  volume  = {237},
  pages   = {447--449},
  year    = {1972},
  doi     = {10.1038/237447a0}
}

@article{Nurmia1964,
  author  = {Nurmia, M. and Graeffe, G. and Valli, K. and Aaltonen, J.},
  title   = {Alpha activity of {Sm}-146},
  journal = {Ann. Acad. Sci. Fenn. Ser. A VI},
  volume  = {148},
  pages   = {1},
  year    = {1964}
}

@article{Chabanat1998,
  author  = {Chabanat, E. and Bonche, P. and Haensel, P. and Meyer, J. and Schaeffer, R.},
  title   = {A Skyrme parametrization from subnuclear to neutron star densities Part II. Nuclei far from stabilities},
  journal = {Nucl. Phys. A},
  volume  = {635},
  pages   = {231},
  year    = {1998},
  doi     = {10.1016/S0375-9474(98)00180-8}
}

@article{Tang2025,
  author  = {Tang, Shulin and Qian, Yibin and Liu, Wei and Ren, Zhongzhou},
  title   = {Examining evidence for a shorter $^{146}$Sm--$^{142}$Nd chronology in the early solar system},
  journal = {Phys. Rev. C},
  volume  = {111},
  pages   = {L052801},
  year    = {2025},
  doi     = {10.1103/PhysRevC.111.L052801}
}

@article{Bartel1982,
  author  = {Bartel, J. and Quentin, P. and Brack, M. and Guet, C. and H{\aa}kansson, H.-B.},
  title   = {Towards a better parametrisation of Skyrme-like effective forces: A critical study of the SkM force},
  journal = {Nucl. Phys. A},
  volume  = {386},
  pages   = {79},
  year    = {1982},
  doi     = {10.1016/0375-9474(82)90403-1}
}

@article{Dobaczewski1984,
  author  = {Dobaczewski, J. and Flocard, H. and Treiner, J.},
  title   = {Hartree-Fock-Bogolyubov description of nuclei near the neutron drip line},
  journal = {Nucl. Phys. A},
  volume  = {422},
  pages   = {103},
  year    = {1984},
  doi     = {10.1016/0375-9474(84)90433-0}
}

@article{Lane1958,
  author  = {Lane, A. M. and Thomas, R. G.},
  title   = {R-Matrix Theory of Nuclear Reactions},
  journal = {Rev. Mod. Phys.},
  volume  = {30},
  pages   = {257},
  year    = {1958},
  doi     = {10.1103/RevModPhys.30.257}
}

@article{Satula1998,
  author  = {Satu{\l}a, W. and Dobaczewski, J. and Nazarewicz, W.},
  title   = {Odd-Even Staggering of Nuclear Masses: Pairing or Shape Effect?},
  journal = {Phys. Rev. Lett.},
  volume  = {81},
  pages   = {3599},
  year    = {1998},
  doi     = {10.1103/PhysRevLett.81.3599}
}

@book{RingSchuck,
  author    = {Ring, P. and Schuck, P.},
  title     = {The Nuclear Many-Body Problem},
  publisher = {Springer-Verlag},
  address   = {New York},
  year      = {1980}
}

@article{Ni2008,
  author  = {Ni, D. and Ren, Z. and Dong, T. and Xu, C.},
  title   = {Unified formula of half-lives for $\alpha$ decay and cluster radioactivity},
  journal = {Phys. Rev. C},
  volume  = {78},
  pages   = {044310},
  year    = {2008},
  doi     = {10.1103/PhysRevC.78.044310}
}

@article{Royer2000,
  author  = {Royer, G.},
  title   = {Alpha emission and spontaneous fission through quasi-molecular shapes},
  journal = {J. Phys. G},
  volume  = {26},
  pages   = {1149},
  year    = {2000},
  doi     = {10.1088/0954-3899/26/8/305}
}

@article{Marks2014,
  author  = {Marks, N. E. and Borg, L. E. and Hutcheon, I. D. and Jacobsen, B. and Clayton, R. N.},
  title   = {Samarium--neodymium chronology and rubidium--strontium systematics of an {Allende} calcium--aluminum-rich inclusion with implications for $^{146}${Sm} half-life},
  journal = {Earth Planet. Sci. Lett.},
  volume  = {405},
  pages   = {15--24},
  year    = {2014},
  doi     = {10.1016/j.epsl.2014.08.017}
}

@article{Boyet2010,
  author  = {Boyet, M. and Carlson, R. W. and Horan, M.},
  title   = {Old {Sm}--{Nd} ages for cumulate eucrites and redetermination of the solar system initial $^{146}${Sm}/$^{144}${Sm} ratio},
  journal = {Earth Planet. Sci. Lett.},
  volume  = {291},
  pages   = {172--181},
  year    = {2010},
  doi     = {10.1016/j.epsl.2010.01.010}
}

@article{Caro2011,
  author  = {Caro, G.},
  title   = {Early Silicate {Earth} Differentiation},
  journal = {Annu. Rev. Earth Planet. Sci.},
  volume  = {39},
  pages   = {31--58},
  year    = {2011},
  doi     = {10.1146/annurev-earth-040610-133400}
}

@article{ONeil2008,
  author  = {O'Neil, J. and Carlson, R. W. and Francis, D. and Stevenson, R. K.},
  title   = {Neodymium-142 evidence for {Hadean} mafic crust},
  journal = {Science},
  volume  = {321},
  pages   = {1828--1831},
  year    = {2008},
  doi     = {10.1126/science.1161925}
}

@article{Rankenburg2006,
  author  = {Rankenburg, K. and Brandon, A. D. and Neal, C. R.},
  title   = {Neodymium isotope evidence for a chondritic composition of the {Moon}},
  journal = {Science},
  volume  = {312},
  pages   = {1369--1372},
  year    = {2006},
  doi     = {10.1126/science.1126114}
}

@article{Borg2016,
  author  = {Borg, L. E. and Brennecka, G. A. and Symes, S. J. K.},
  title   = {Accretion timescale and impact history of {Mars} deduced from the isotopic systematics of martian meteorites},
  journal = {Geochim. Cosmochim. Acta},
  volume  = {175},
  pages   = {150--167},
  year    = {2016},
  doi     = {10.1016/j.gca.2015.12.002}
}

@article{QiXu2009,
  author  = {Qi, C. and Xu, F. R. and Liotta, R. J. and Wyss, R. and Zhang, M. Y. and Asawatangtrakuldee, C. and Hu, D.},
  title   = {Microscopic mechanism of charged-particle radioactivity and generalization of the {Geiger-Nuttall} law},
  journal = {Phys. Rev. C},
  volume  = {80},
  pages   = {044326},
  year    = {2009},
  doi     = {10.1103/PhysRevC.80.044326}
}

@article{QiLiotta2019,
  author  = {Qi, C. and Liotta, R. and Wyss, R.},
  title   = {Recent developments in radioactive charged-particle emissions and related phenomena},
  journal = {Prog. Part. Nucl. Phys.},
  volume  = {105},
  pages   = {214--251},
  year    = {2019},
  doi     = {10.1016/j.ppnp.2018.11.003}
}

@article{Akrawy2017,
  author  = {Akrawy, D. T. and Poenaru, D. N.},
  title   = {Alpha decay calculations with a new formula},
  journal = {J. Phys. G},
  volume  = {44},
  pages   = {105105},
  year    = {2017},
  doi     = {10.1088/1361-6471/aa8527}
}

@article{Akrawy2019,
  author  = {Akrawy, D. T. and Hassanabadi, H. and Hosseini, S. S. and Santhosh, K. P.},
  title   = {Systematic study of $\alpha$ decay half-lives using new universal decay law},
  journal = {Int. J. Mod. Phys. E},
  volume  = {28},
  pages   = {1950075},
  year    = {2019},
  doi     = {10.1142/S0218301319500757}
}

@article{RenRen2012,
  author  = {Ren, Yuejiao and Ren, Zhongzhou},
  title   = {New {Geiger-Nuttall} law for $\alpha$ decay of heavy nuclei},
  journal = {Phys. Rev. C},
  volume  = {85},
  pages   = {044608},
  year    = {2012},
  doi     = {10.1103/PhysRevC.85.044608}
}

@article{Bouvier2010,
  author  = {Bouvier, A. and Wadhwa, M.},
  title   = {The age of the {Solar} {System} redefined by the oldest {Pb}--{Pb} age of a meteoritic inclusion},
  journal = {Nat. Geosci.},
  volume  = {3},
  pages   = {637--641},
  year    = {2010},
  doi     = {10.1038/ngeo941}
}

@misc{SM,
  note = {See Supplemental Material at [URL]
          for further details regarding the quartet formation amplitude
          (Sec.~S-I), the interaction potential and preformation probability
          from the shell--cluster coupling (Sec.~S-II), the calculated and experimental half-lives of the
          neighboring emitters (Sec.~S-III), and the inversion of the
          $^{146}$Sm--$^{142}$Nd chronometer (Sec.~S-IV), which also
          includes Refs.~\cite{RingSchuck,Soloviev1962,MangRasmussen1962,Mang1964,Macfarlane1961,AlBataina1987,Gupta1970,Casali2016,Nica2026,Prestwood1981,Fulop2003,MacFarlane1961Dy,Holden1985,Mahunka1966,Golovkov1968,Gono1971,Chiera2022,Nica2025,NUBASE2020,Villa2020}}
}

@article{Amelin2025,
  author  = {Amelin, Yuri and Yin, Qing-Zhu},
  title   = {Recent progress and future prospects of the early solar system chronology},
  journal = {Natl. Sci. Rev.},
  volume  = {12},
  number  = {9},
  pages   = {nwaf281},
  year    = {2025},
  doi     = {10.1093/nsr/nwaf281}
}

@article{Andreyev2013,
  author  = {Andreyev, A. N. and Huyse, M. and Van Duppen, P. and Qi, C. and
             Liotta, R. J. and Antalic, S. and Ackermann, D. and Franchoo, S. and
             He{\ss}berger, F. P. and Hofmann, S. and Kojouharov, I. and
             Kindler, B. and Kuusiniemi, P. and Lesher, S. R. and Lommel, B. and
             Mann, R. and Nishio, K. and Page, R. D. and Streicher, B. and
             {\v S}{\'a}ro, {\v S}. and Sulignano, B. and Wiseman, D. and Wyss, R. A.},
  title   = {Signatures of the $Z=82$ Shell Closure in $\alpha$-Decay Process},
  journal = {Phys. Rev. Lett.},
  volume  = {110},
  pages   = {242502},
  year    = {2013},
  doi     = {10.1103/PhysRevLett.110.242502}
}

@article{QiAndreyev2010,
  author  = {Qi, C. and Andreyev, A. N. and Huyse, M. and Liotta, R. J. and
             Van Duppen, P. and Wyss, R. A.},
  title   = {Abrupt changes in $\alpha$-decay systematics as a manifestation of
             collective nuclear modes},
  journal = {Phys. Rev. C},
  volume  = {81},
  pages   = {064319},
  year    = {2010},
  doi     = {10.1103/PhysRevC.81.064319}
}

@article{BennaceurDobaczewski2005,
  author  = {Bennaceur, K. and Dobaczewski, J.},
  title   = {Coordinate-space solution of the {S}kyrme--{H}artree--{F}ock--{B}ogolyubov
             equations within spherical symmetry. The program {HFBRAD} (v1.00)},
  journal = {Comput. Phys. Commun.},
  volume  = {168},
  pages   = {96},
  year    = {2005},
  doi     = {10.1016/j.cpc.2005.02.002}
}

@article{Soloviev1962,
  author  = {Soloviev, V. G.},
  title   = {Effect of pairing correlations on the alpha decay rates},
  journal = {Phys. Lett.},
  volume  = {1},
  pages   = {202},
  year    = {1962},
  doi     = {10.1016/0031-9163(62)90354-2}
}

@article{MangRasmussen1962,
  author  = {Mang, H. J. and Rasmussen, J. O.},
  title   = {Shell-model calculations of alpha decay rates of even-even spheroidal nuclei},
  journal = {Mat. Fys. Skr. Dan. Vid. Selsk.},
  volume  = {2},
  number  = {3},
  pages   = {No.~3},
  year    = {1962}
}

@article{Mang1964,
  author  = {Mang, H. J.},
  title   = {Alpha Decay},
  journal = {Annu. Rev. Nucl. Sci.},
  volume  = {14},
  pages   = {1},
  year    = {1964},
  doi     = {10.1146/annurev.ns.14.120164.000245}
}

@article{Shollenberger2026,
  author  = {Shollenberger, Q. R. and Borg, L. E. and Kavner, A. R. L. and Kmak, K. N. and Boyd, S. T. P. and Despotopulos, J. D. and Drury, O. B. and Jovanovic, I. and Friedrich, S. and Gallant, A. T. and Kim, I. W. and Kim, Y.-H. and Kunz, P. and Kwiatkowski, A. A. and Kwon, D. H. and Lee, D. and Scielzo, N. D. and Kim, G.-B.},
  title   = {A new half-life for $^{146}${Sm}: {Implications} for lunar chronology},
  journal = {Earth Planet. Sci. Lett.},
  volume  = {695},
  pages   = {120326},
  year    = {2026},
  doi     = {10.1016/j.epsl.2026.120326}
}

@article{Macfarlane1961,
  author  = {Macfarlane, R. D. and Kohman, T. P.},
  title   = {Natural alpha radioactivity in medium-heavy elements},
  journal = {Phys. Rev.},
  volume  = {121},
  pages   = {1758},
  year    = {1961},
  doi     = {10.1103/PhysRev.121.1758}
}

@article{AlBataina1987,
  author  = {Al-Bataina, B. and J\"anecke, J.},
  title   = {Half-lives of long-lived alpha emitters},
  journal = {Radiochim. Acta},
  volume  = {42},
  pages   = {159},
  year    = {1987},
  doi     = {10.1524/ract.1987.42.4.159}
}

@article{Gupta1970,
  author  = {Gupta, M. C. and MacFarlane, R. D.},
  title   = {The natural alpha radioactivity of samarium},
  journal = {J. Inorg. Nucl. Chem.},
  volume  = {32},
  pages   = {3425},
  year    = {1970},
  doi     = {10.1016/0022-1902(70)80149-X}
}

@article{Casali2016,
  author  = {Casali, N. and Dubovik, A. and Nagorny, S. and Nisi, S. and Orio, F. and Pattavina, L. and Pirro, S. and Sch\"affner, K. and Tupitsyna, I. and Yakubovskaya, A.},
  title   = {Cryogenic detectors for rare alpha decay search: A new approach},
  journal = {J. Low Temp. Phys.},
  volume  = {184},
  pages   = {952},
  year    = {2016},
  doi     = {10.1007/s10909-016-1507-y}
}

@article{Nica2026,
  author  = {Nica, N.},
  title   = {Nuclear Data Sheets for {A} = 148},
  journal = {Nucl. Data Sheets},
  volume  = {208},
  pages   = {1},
  year    = {2026},
  doi     = {10.1016/j.nds.2026.01.001}
}

@article{Prestwood1981,
  author  = {Prestwood, R. J. and Curtis, D. B. and Cappis, J. H.},
  title   = {Half-life of $^{148}${Gd}},
  journal = {Phys. Rev. C},
  volume  = {24},
  pages   = {1346},
  year    = {1981},
  doi     = {10.1103/PhysRevC.24.1346}
}

@article{Fulop2003,
  author  = {F\"ul\"op, {\relax Zs}. and Bartha, L. and Gy\"urky, {\relax Gy}. and Somorjai, E. and Kubono, S. and Kudo, H. and Kaji, D.},
  title   = {The half-life of $^{148}${Gd}},
  journal = {Nucl. Phys. A},
  volume  = {718},
  pages   = {688},
  year    = {2003},
  doi     = {10.1016/S0375-9474(03)00889-3}
}

@article{MacFarlane1961Dy,
  author  = {MacFarlane, R. D.},
  title   = {Dysprosium-154, a long-lived $\alpha$-emitter},
  journal = {J. Inorg. Nucl. Chem.},
  volume  = {19},
  pages   = {9},
  year    = {1961},
  doi     = {10.1016/0022-1902(61)80039-0}
}

@techreport{Holden1985,
  author      = {Holden, N. E.},
  title       = {Long-lived heavy mass elements half-lives ({A} $>$ 125)},
  institution = {Brookhaven National Laboratory},
  number      = {BNL-NCS-36960},
  year        = {1985}
}

@article{Mahunka1966,
  author  = {Mahunka, I. and F\'enyes, T.},
  title   = {Investigation of the alpha spectrum of {Dy} isotopes},
  journal = {Bull. Acad. Sci. USSR, Phys. Ser.},
  volume  = {29},
  pages   = {1126},
  year    = {1966}
}

@article{Golovkov1968,
  author  = {Golovkov, N. A. and Gromov, K. {\relax Ya}. and Lebedev, N. A. and Makhmudov, B. and Rudnev, A. S. and Chumin, V. G.},
  title   = {Concerning alpha decay of {Dy}, {Tb}, {Gd} and {Eu} isotopes},
  journal = {Bull. Acad. Sci. USSR, Phys. Ser.},
  volume  = {31},
  pages   = {1657},
  year    = {1968}
}

@article{Gono1971,
  author  = {G\=on\=o, Y. and Hiruta, K.},
  title   = {Non-existence of isomeric $\alpha$-decay in {Dy}$^{154}$},
  journal = {J. Phys. Soc. Jpn.},
  volume  = {30},
  pages   = {1241},
  year    = {1971},
  doi     = {10.1143/JPSJ.30.1241}
}

@article{Chiera2022,
  author  = {Chiera, N. M. and Dressler, R. and Sprung, P. and Talip, Z. and Schumann, D.},
  title   = {High precision half-life measurement of the extinct radio-lanthanide {Dysprosium-154}},
  journal = {Sci. Rep.},
  volume  = {12},
  pages   = {8988},
  year    = {2022},
  doi     = {10.1038/s41598-022-12684-6}
}

@article{Nica2025,
  author  = {Nica, N.},
  title   = {Nuclear Data Sheets for {A} = 154},
  journal = {Nucl. Data Sheets},
  volume  = {200},
  pages   = {2},
  year    = {2025},
  doi     = {10.1016/j.nds.2025.02.002}
}

\end{document}